\documentclass{article}
\usepackage{epsf}
\usepackage{complex-systems}

\begin{document}

\title{Measuring Mutual Information in Random\\
Boolean Networks} 
\author{\authname{Bartolo Luque}\\[2pt]  
\authadd{Centro de Astrobiolog\'{\i}a (CAB), Ciencias del Espacio, INTA,}\\ 
\authadd{Carretera de Ajalvir km. 4, 28850 Torrej\'on de Ardoz, Madrid, Spain}%
\and
\authname{Antonio Ferrera}\\[2pt]
\authadd{Centro de Astrobiolog\'{\i}a (CAB), Ciencias del Espacio, INTA,}\\ 
\authadd{Carretera de Ajalvir km. 4, 28850 Torrej\'on de Ardoz, Madrid, Spain}
} 
\markboth{B. Luque and A. Ferrera} 
{Mutual Information Measures in Random Boolean Networks} 
\maketitle

\begin{abstract} 
During the last few years an area of active research in the field of complex systems is 
that of their information storing and processing abilities. Common opinion has it that the most 
interesting beaviour of these systems is found ``at the edge of chaos'', which would seem 
to suggest that complex systems may have inherently non-trivial information proccesing 
abilities in the vicinity of sharp phase transitions. A comprenhensive, quantitative 
understanding of why this is the case is however still lacking. Indeed, even ``experimental'' 
(i.e., often numerical) evidence that this is so has been questioned for a number of systems. 
In this paper we will investigate, both numerically and analitically, the behavior of Random 
Boolean Networks (RBN's) as they undergo their order-disorder phase transition. We will 
use a simple mean field approximation to treat the problem, and without lack 
of generality we will concentrate on a particular value for the connectivity of the system. 
In spite of the simplicity of our arguments, 
we will be able to reproduce analitically the amount of mutual information contained in the system 
as measured from numerical simulations.

\end{abstract}

\section{Introduction} 
The amount of information that a system is able to process (and/or store) plays an essential role
when one tries to quantify the level of ``complexity'' of a system, and indeed often the mutual 
information [1] stored in the system (or a concept derived from it, such as the past-future mutual 
information) is used as a measure of its statistical complexity [2].

Over the last decade a number of authors have carried out work towards understanding 
under what conditions can we expect to maximize the information procesing capabilities of different 
types of complex systems. For instance, Langton and others [3,4] investigated the behavior of 
Cellular Automata (CA), while Crutchfield, Young and others [5] have been concerned 
mainly with iterated function systems and computational complexity in this area. The 
definitions used for complexity were rather problem dependent, and not surprisingly two main 
approaches to measuring statistical complexity have been developed over the years, as well as a large
number of other ``ad hoc'' methods for describing structure. The first line of work uses information 
theory [6-9], whereas the second approach defines complexity using computation theoretic tools [5,10].

In spite of this model dependence, the common picture that seemed to emerge from this work was 
that complex systems were able to show a maximally varied and self-organizative behaviour (i.e., 
maximally complex behaviour) in the vicinity of sharp phase transitions [11]. Since these transitions 
often belonged to the class commonly known in statistical mechanics as order-disorder phase 
transitions, this naturally led to the notion that maximally interesting behaviour of complex 
systems takes place ``at the edge of chaos'', in an expression coined by Langton [3]. (Note however 
that the disordered phase does not neccesarily need to be chaotic in the strict sense of the word, 
i.e., ergodyc.) The underlying reason was simple and appeling enough, neither very ordered systems 
with static structures, nor disordered systems in which information can not be persistently stored 
are capable of complex information processing tasks.

The actual verification of the fact that the mutual information (or definitions of statistical 
complexity based on other approaches) had a maximum in the vicinity 
of the relevant phase transitions were a trickier business though. Early results by Langton for 
CA's [3,4] and by Crutchfield [5,10] for iterated dynamics showing sharp peaks in 
complexity as a function of the degree of order in the system at what appeared to be 
phase transitions were subsequently shown to be critically dependent on the particular 
measure of order choosen [2]. After this, Arnold [12] showed numerically that the 2-dimensional 
Ising model indeed had a maximum of statistical complexity (defined through past-future mutual 
information) at its order-disorder transition. 

Without wanting to go into the debate of what exactly constitutes a good measure of complexity -a 
debate often riddled with the specifics of the particular problem at hand-, it would seem clear 
though that complexity and information must bear a close relationship. We will thus concern 
ourselves in this paper with the mutual information contained in Random Boolean Networks (RBN) [13] 
and its behavior as the networks undergo their order-disorder phase transition (for a view point 
computational see [18]). By using a mean field approximation and assuming Markovian behaviour of
 the automata, we will show both 
numerically and analytically that the mutual information stored in the network indeed has a maximum at 
the transition point. 

\section{Random Boolean Networks} 
Random Boolean Networks (RBN) [13] are systems composed of a number N of automata ($i=1,...,N$) 
with only two states 
available (say $0$ and $1$ for instance), each having associated a Boolean function $f_i$ of $K$
Boolean arguments that 
will be used to update the automaton state at each time step. Each automaton $i$ will then 
have associated $K$ other automata $i_{1}, i_{2}, ..., i_{k}$ (the inputs or vicinity 
of $i$), 
whose states $(x_{i_{1}}, x_{i_{2}}, ..., x_{i_{k}})$ will be the entries of $f_{i}$. 
That is, the automaton $i$ will change its state $x_{i}$ at each time step according to the 
rule
$$x_{i}(t+1)=f_{i}(x_{i_{1}}(t), x_{i_{2}}(t), ..., x_{i_{k}}(t)). \eqno (1)$$

Both $f_{i}$ and the identity of its $K$ inputs are initially assigned to the 
automaton $i$ at random. (In particular, the $N$ $f$'s are created by randomly generating 
outputs of value one with a probability $p$, and of value zero with a probability $1-p$, 
where $p$ is called the bias of the network). This initial assignation will be maintained 
throught the evolution of the system, so we will be dealing with a quenched system.  
Even keeping this assignation fixed, the number of possible networks that we can form for 
given values of $N$ and $K$ is extraordinarily high (a total of $(2^{2^{K}}N^{K})^{N}$ 
possible networks). Thus, if we want to study general characteristics of RBN systems we are 
inevitably led to an statistical approach.

One fact that can be observed for all RBN's is that although the number of available states 
for a network of size $N$ grows like $2^{N}$, the dynamics of 
the net separates the possible states into disjoint sets, attractor basins. Each basin will 
lead the system to a different attractor. However, since the number of states 
available is finite and the quenched system is 
fully deterministic, we can be sure that the system will at some point retrace its 
steps in the form of periodic cycles. Thus attractors will neccesarily be periodic sets of states. 
Since after a transient any initial state will end up in one attractor or another, their period (or 
rather their average period) will set the typical time scale characterizing an RBN. 

It has been known for some time now [13] that RBN's show two different phases separated, 
for a given value of $p$, by a critical value of $K$, $K_c$:
\enumerate{
\item an ordered phase for $K<K_c$ in 
wich the networks cristalize in a pattern after a short transient. In this phase almost all of 
the automata remain in a completely frozen state and the average period $<T>$ of the attractors 
scale with $N$ as a power law and 
\item  a disordered phase for $K>K_c$. All patterns are lost and 
the automata appear to be 
in a completely disordered state, switching from one state to another seemingly at random. The 
period of the attractors become unobservable in practice because $<T>$ grows exponentially with 
$N$, thereby rendering the system free of any time scale [14].
}
\endenumerate
This behaviour naturally induced the 
conjecture that at $K_c$ the RBN's undergo a second order phase transition. This conjecture has 
been prooven correct and some more information about the transition has been 
gained [15]. For instance, as we change 
the value of $p$ the critical value $K_{c}$ at which the transtion takes place also changes and 
a ``critical line'' appears, as shown in  Figure 1. As was shown by [16]
 this line corresponds to

$$ K={1\over 2p(1-p)}. \eqno(2)$$

In the insets of Figure 1, three sets of states of a network with $N=50$ and 
$K=3$ are also shown as we move from the disordered state to the ordered one by changing $p$, 
showing a typical order-disorder transition. Each set of states contains 50 consecutive states, time running upwards along 
the vertical axys.

\begin{figure}
\vspace{8cm}
\includegraphics{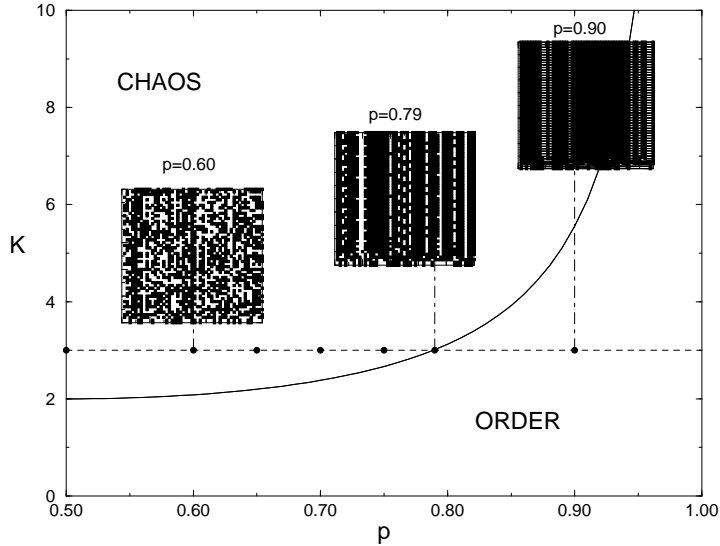}
\label{criticalline-fig}
\caption{The boundary between the chaotic and the ordered phase is shown in a K-p phase diagram. For a constant value of K, 
K=3, three set examples of a $N=50$ network are shown for $p=0.60$ (disordered phase), $p=0.79$ (over the critical 
line), and $p=0.90$ (ordered phase). Each run contains 50 consecutive states, time increasing upwards along 
the vertical axys. 
}
\end{figure}   

\section{Self-overlap in RBN}

Since RBN's appear undergo an order-disorder phase transition, a useful way to caharacterize the state of the 
system will be its ``self-overlap'' $a$. This is simply defined to be one minus the Hamming distance between 
an automaton at time $t$ and itself at time $t+1$, averaged over all automata and times. Let us expand on this. 

Let us suppose that we generate an RBN with bias $p$, and a random initial condition. We let the system evolve 
until the transient dies out and we are inside an attractor cycle, and then compute the 
states of the system for a number of time steps equal to the number of automata in the system (that is, from 
$t=1$ to $t=10,000$ for the $N=10,000$ network that we have used. Each experimental computer point 
in all figures is the average of $100$ differents networks with random initial conditions).
Let us suppose that we count the number of times that an automata is in the state $1$ both at time 
$t$ and $t+1$, and average over all automata and time steps. This will give us the ``1 state
self-overlap'', $a_{11}$. Repeating this procedure with the $0$ state will then obviously give us 
the ``zero state self-overlap'', $a_{00}$. Then, $a$ will simply be given by
$$a=a_{11}+a_{00}. \eqno(3)$$
On the other hand, we can analogously define $a_{10}$ and $a_{01}$. Note that by symmetry we have 
to have  $a_{10}=a_{01}$ even with $p\neq1/2$, since $a_{10}$ and $a_{10}$ are the joint 
probability distributions, not the conditional probabilities of transitioning from $1$ to $0$ or 
viceversa.
  
It is then fairly easy to find the equation that describes the evolution of $a$. If we define $P$ to be
$$P=p^2+(1-p)^2, \eqno(4)$$
then it is not difficult to convince oneself that in a mean field approximation we must have 
$$ a_{t+1} = a_t^K+P(1-a_t^K),  \eqno(5)$$
where $K$ is the connectivity of the net.
This equation forces $a$ to evolve towards fixed points, $a_t\rightarrow a^*$, that will 
depend on $K$ and $p$. The stability analysis of (5) for $a^*=1$ gives the critical line
(2) separating the ordered phase ($a^*=1$) from the disordered phase ($a^*< 1$).
This is shown in Figure 2, where the evolution of $a$ 
given by (5) (solid line) is plotted against the results of the numerical simulations (dots). 
The evolution lasts for as long as 
it takes the transient to die out, and once the system is in the attractor cycle $a$ takes on its 
fixed point value (from now on we drop the star and designate the fixed point value simply by $a$).

\begin{figure}
\vspace{8cm}
\includegraphics{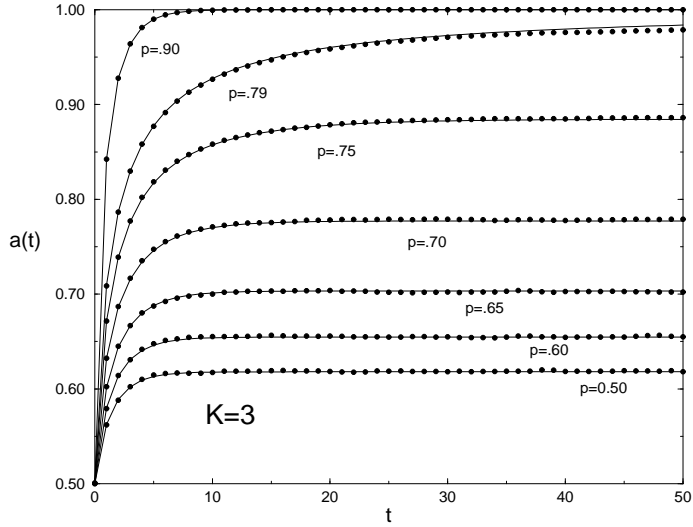}
\label{auto-fig}
\caption{Plot of $a(t)$ vs. $t$ showing the numerical results from the simulations (dots) and the values predicted by the 
mean field approximation (5), for $K=3$. Different values of the bias $p$, from $p=0.5$ to $p=0.9$ with $p_c=0.79$ are shown.
}
\end{figure} 

\begin{figure}
\vspace{8cm}
\label{autos-fig}
\includegraphics{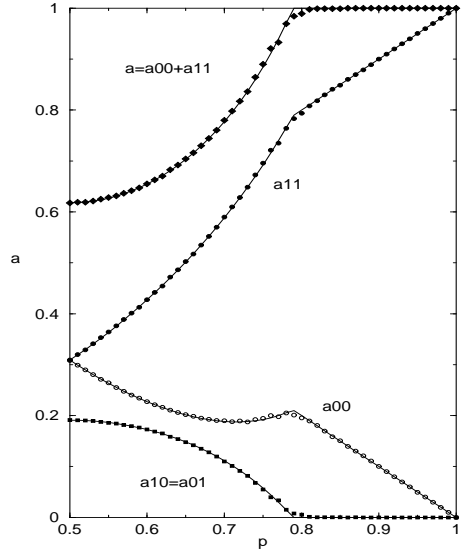}
\caption{Analytical (continous line) and numerical (dots) results for the $a_{\alpha \beta}$ as $p$ ranges from 0.5 to 1
}
\end{figure}

Let us now obtain analitycal expressions for the $a_{\alpha \beta}$ from our knowledge of $a$, the 
normalization conditions and the fact that $a_{10}=a_{01}$. By definition (3) and by normalization
$$a_{00}+a_{01}+a_{10}+a_{11}=1, \eqno(6)$$
so that
$$a_{01}+a_{10}=1-a. \eqno (7)$$
But then, by symmetry,
$$a_{10}=a_{01}={1-a\over2}. \eqno (8)$$
We still have two more normalization conditions, derived from the fact that the probability of 
finding a mean field automaton in the state 1 is $p$, and $1-p$ for the state $0$
$$a_{11}+a_{10}=p, \eqno (9)$$
$$a_{00}+a_{01}=1-p, \eqno (10)$$
whence
$$a_{00}={a\over2}-\Biggl(p-{1\over2}\Biggr),$$
$$a_{11}={a\over2}+\Biggl (p-{1\over2}\Biggr),$$
which satisfy (3) above. Figure\ref{autos-fig} shows the analytical expressions for the $a_{\alpha 
\beta}$ (solid lines) together with the results from the numerical simulations (dots). 

So far we have simply approximated the whole network by a set of mean field automata. However, since the 
$a_{\alpha \beta}$ are equivalent to $p_{\alpha \cap \beta}$ we can now calculate the 
conditional probabilities
$$p_{\alpha \mid \beta}={a_{\alpha \beta}\over p_{\beta}} \eqno (13).$$
If we now assume that our mean field automata are Markovian these conditional probabilities will 
completely characterize their transition probabilities [1]. Therefore, the transition matrix for the mean 
field Markovian automaton is:
$${\bf T}=\pmatrix {p_{0 \mid 0}&p_{1 \mid 0}\cr p_{0 \mid 1}&p_{1 \mid 1}\cr}=
\pmatrix {{a+1-2p \over 2(1-p)}&{1-a \over 2(1-p)}\cr {1-a \over 2p}& {a-1+2p \over 2p}\cr} \eqno(14)$$
which satisfy
$$\sum_{\alpha=0}^1{p_{\alpha \mid 0}}=1, \eqno(15)$$
$$\sum_{\alpha=0}^1{p_{\alpha \mid 1}}=1, \eqno(16)$$
where $p_{\alpha \mid 0}, p_{\alpha \mid 1}$ are the probabilities of transitioning from the 
states $0,1$ to the state $\alpha$.

Thus, we have now reduced the whole network to a set of mean field automata evolving independently under 
Markovian conditions, all the effects of their interactions being encoded in $a$. To compute the 
past-future mutual information stored in the system we only have to apply information theory [2,17]. The 
one-automaton entropy is simply
$$H(x_{t+1})=-p\log{p}-(1-p)\log{(1-p)}, \eqno(17)$$
whereas the Shannon uncertainty associated to the Markovian evolution of this automaton will be
$$H(x_{t+1} \mid x_{t})=p H(x_{t+1} \mid x_t=1)+ (1-p) H(x_{t+1} \mid x_t=0), \eqno(18)$$
with
$$H(x_{t+1} \mid x_t=1)=-{a-1+2p \over 2p}\log{\Biggr({a-1+2p \over 2p}\Biggl)}-$$ 
$$- {1-a \over 2p}\log{\Biggr( {1-a \over 2p}\Biggl)}, \eqno(19)$$
and
$$H(x_{t+1} \mid x_t=0)=-{a+1-2p \over 2(1-p)}\log{\Biggr({a+1-2p \over 2(1-p)}\Biggl)}-$$ 
$$- {1-a \over 2(1-p)}\log{\Biggr( {1-a \over 2(1-p)}\Biggl)}. \eqno(20)$$
The uncertainty is thus,
$$H(x_{t+1} \mid x_{t})=-{a-1+2p \over 2}\log{\Biggr({a-1+2p \over 2p}\Biggl)}-$$ 
$$-{1-a \over p}\log{\Biggr( {1-a \over 2p}\Biggl)}-{a+1-2p \over 2}\log{\Biggr({a+1-2p \over 2(1-p)}\Biggl)}-$$ 
$$- {1-a \over 2}\log{\Biggr( {1-a \over 2(1-p)}\Biggl)}, \eqno(21)$$
whence the past-future mutual information will be:
$$I=H(x_{t+1})-H(x_{t+1} \mid x_{t})=$$
$$=-p\log{p}-(1-p)\log{(1-p)} +{a-1+2p \over 2}\log{\Biggr({a-1+2p \over 2p}\Biggl)} +$$
$$+{1-a \over p}\log{\Biggr( {1-a \over 2p}\Biggl)}+{a+1-2p \over 2}\log{\Biggr({a+1-2p \over 2(1-p)}\Biggl)}+$$ 
$$+{1-a \over 2}\log{\Biggr( {1-a \over 2(1-p)}\Biggl)}. \eqno(22)$$
Figure 4 shows the analytical expressions (solid lines) as well the experimental results 
from the simulations 
for both the one-automaton entropy (dots) and the Shannon uncertainty (triangles). Note how the uncertainty 
is always smaller and decays faster than the one-automaton entropy. In particular, for $p \ge p_c$ 
(where $p_c= 0.79$ for our net with $K=3$), we have $a=1$ and $H(x_{t+1} \mid x_{t})=0$. Thus in the ordered 
phase the mutual information becomes simply the one-automaton entropy. Given  this discussion, it is obvious 
that the mutual information that can be stored in the system has to have a maximum precisely at $p_c$. This 
is shown in Figure 4, where the mutual information $I$ is plotted against $p$ (again, 
both the analytical expression above as well as the experimental results).

\begin{figure}
\vspace{8cm}
\includegraphics{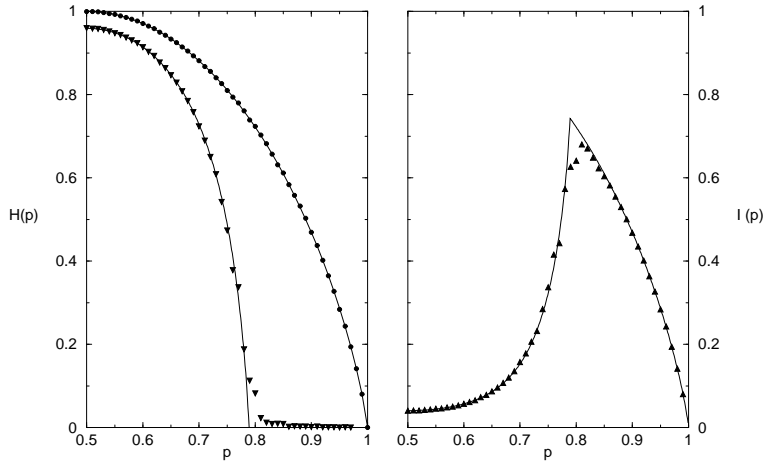}
\label{informacion-fig}
\caption{Numerical and analytical results for both $H(x_{t+1})$ (filled dots for the numerical results) and 
$H(x_{t+1} \mid x_{t})$ (filled triangles) are shown in the left hand side figure. Note how $H(x_{t+1} \mid x_{t})$ is 
always smaller and decays faster than $H(x_{t+1})$, becoming zero in the ordered phase. In the right hand side figure 
$I$ vs. $p$ is shown, showing a peak at the critical value $p_c=0.79$ as expected. 
}
\end{figure}

Finally, in Figure 5 the mutual information is plotted against the one-automaton entropy $H$. 
From $H=0$ corresponding to $p=1$ to $H \approx 0.75$ which corresponds to the critical value $p=p_c$, we see 
that $I$ is just a straight line of slope $1$. This is as it should be, since as we just saw 
$H(x_{t+1} \mid x_{t})$ is zero
for $p$ beyond $p_c$, and $I = H(x_{t+1})$ in this region. Precisely at $H(p_c)$, $I$ reaches a maximum, and 
beyond this point it starts to decay non-lineary as the Shanon uncertainty switches on.  
\begin{figure}
\vspace{8cm}
\includegraphics{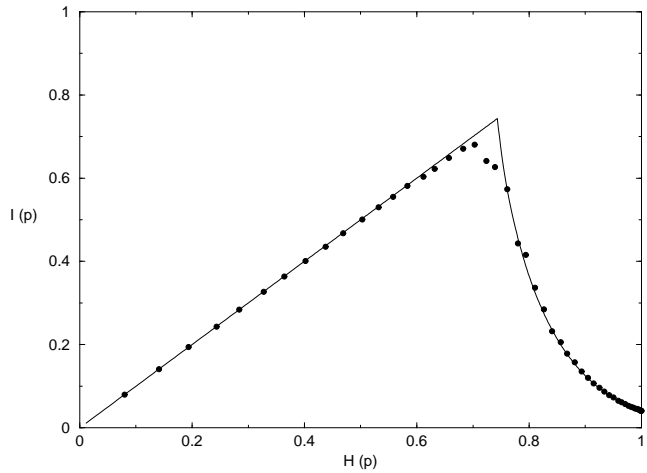}
\label{HversusI-fig}
\caption{The past-future mutal information $I$ vs. the one-automaton entropy $H(x_{t+1})$. From $H=0$ (corresponding 
to $p=1$) to $H(p_c)$ we have $I = H(x_{t+1})$ since $H(x_{t+1} \mid x_{t})=0$. Therefore, in this region the mutual 
information simply increases linearly with the one block entropy. Beyond this point however we enter the disordered phase 
and $H(x_{t+1} \mid x_{t})$ switches on, growing faster (in absolute value) than the one block entropy. Therefore, $I$ 
shows a peak exactly at the transition point.   
}
\end{figure}

\section{Conclusions}

By using a mean field approximation and a Markovian ansatz for the evolution
of an RBN, we have been able to show with a few, back of the envelope type
of calculations, that the past-future mutual information contained in a RBN
reaches a maximum at the point at which this system undergoes its
order-disorder phase transition. Also, in Figure 5 we can see how the mutual
information as a function of the amount of disorder present in the system
(the one-automaton entropy) reaches a maximum at the point that corresponds
to the phase transition. 

Similar results obtained in [3,4] (for CA's) and in [5] (for symbolic
dynamics of the logistic map) were criticized by Li [2] on the ground that
the peak was a artifact created by the particular quantity chosen to measure
the disorder of the system. Thus for instance Li criticizes Langton arguing
that since in the ordered phase we have $I=H(x_t)$, it is only natural for
him to find a straight line as the boundary of his plot of complexity
against disorder (as we do). Li surmises that if instead of using $H(x_t)$
as a measure of the disorder of the system one chooses to use the Shanon
uncertainty of the source $H(x_{t+1}\mid x_t)$ ($H_{t+1|t}$ for short from
now on) the left side of the plot would no longer be a straight line, and
the maximum of $I$ would not be reached for intermediate values of the
disorder. Rather, in the $I.vs.H_{t+1|t}$ plot the maximum of $I$ falls over
the $y$ axis since $I$ reaches a maximum at zero $H_{t+1|t}$, and $I$
monotonically decreases as $H_{t+1|t}$ increases. Thus, the intuitive
picture of the relationship between complexity and disorder proposed by
Langton and others (i.e., unimodal relationship between complexity and
disorder with complexity reaching a maximum at intermediate values of the
latter) would no longer seem to be correct. This Li takes as support to his
conclusion that the dependence of $I$ on the amount of disorder in the
system can take many varied forms. 

We think that the argument just presented, although trivially correct, fails
to capture the essence behind the idea of unimodal dependence between $I$
and the amount of disorder in the system. We should first note that $%
I(H_{t+1|t}=0)$ is not a single valued function. Rather, since $H_{t+1|t}=0$
for $p_c\leq p\leq 1$, at $H_{t+1|t}=0$ $I$ grows from zero (corresponding
to $p=1$) to its maximum value  (corresponding to $p=p_c$). That is, we have
not got rid of the straight line in the $I.vs.H_t$ graph, we have merely
made it into a vertical line placed at $H_{t+1|t}=0$.  Note however that the
maximum of $I$ would still be reached at the transition point between the
two phases of the system. This is in fact the central point of the issue at
hand. The postulated unimodal dependence between $I$ (or complexity) and
disorder rests under the assumption that, as we vary the order parameter,
the system goes from an ordered phase into a disordered one with $I$
attainning its maximum value neither at one phase nor the other, but
precisely at the transition point between them. If the quantity chosen as
the order parameter varies over both phases then $I$ will reach this maximum
for intermediate values of the parameter. If, on the other hand, a whole
phase of the system is mapped into a single value of the order parameter,
then quite obviously the maximum will be at one of the edges of the graph.
Thus one could say that the essence of `unimodality' lyes not on $I$
reaching its maximum for intermediate values of the order parameter, but on
such maximum being at the transition point between the ordered and the
disordered phases.

\label{main-text}

\section*{Acknowledgments}
The authors would like to thank Juan P\'erez Mercader and R. V. Sol\'e for help. This work has been supported
 by the Centro de  Astrobiolog\'{\i}a.

\end{document}